\documentclass[aps,prl,twocolumn,superscriptaddress,floatfix,10pt,letterpaper,showpacs]{revtex4}

\pdfoutput=1

\begin{document}

\begin{titlepage}

\title{Chiral symmetry on the edge of 2D symmetry protected topological phases
}

\author{Xie Chen}
\affiliation{Department of Physics, Massachusetts Institute of
Technology, Cambridge, Massachusetts 02139, USA}

\author{Xiao-Gang Wen}
\affiliation{Perimeter Institute for Theoretical Physics, Waterloo, Ontario, N2L 2Y5 Canada}
\affiliation{Department of Physics, Massachusetts Institute of
Technology, Cambridge, Massachusetts 02139, USA}
\affiliation{Institute for Advanced Study, Tsinghua University,
Beijing, 100084, P. R. China}

\begin{abstract}
Symmetry protected topological (SPT) states are short-range entangled states
with symmetry, which have symmetry protected gapless edge states around a
gapped bulk.  Recently, we proposed a systematic construction of SPT phases in
interacting bosonic systems, however it is not very clear what is the form of
the low energy excitations on the gapless edge. In this paper, we answer this
question for two dimensional bosonic SPT phases with $\mathbb{Z}_N$ and $U(1)$
symmetry.  We find that while the low energy modes of the gapless edges are
non-chiral, symmetry acts on them in a chiral way, i.e. acts on the right
movers and the left movers differently. This special realization of symmetry
protects the gaplessness of the otherwise unstable edge states by prohibiting a
direct scattering between the left and right movers. Moreover, understanding of
the low energy effective theory leads to experimental predictions about the SPT
phases. In particular, we find that \emph{all} the 2D $U(1)$ SPT phases have
\emph{even} integer quantized Hall conductance.

\end{abstract}


\maketitle

\end{titlepage}




\textit{Introduction} -- A recent study shows that gapped quantum  states
belong to two classes: short-range entangled and long-range
entangled.\cite{CGW1038} The long-range entanglement (i.e. the topological
order\cite{W9039}) in the bulk of states is manifested in the existence of gapless
edge modes or degenerate edge sectors.  The short-range entangled states are
trivial and all belong to the same phase if there is no symmetry. However, with
symmetry, even short-range entangled states can belong to different phases.
Those phases are called symmetry protected topological (SPT) phases. The
symmetric short-range entanglement (i.e. the SPT order) is also manifested in
the existence of gapless edge modes around a gapped bulk if the symmetry is not
broken. For example, two and three dimensional topological
insulators\cite{KM0501,BZ0602,KM0502,MB0706,FKM0703,QHZ0824} have a gapped
insulating bulk but host gapless fermion modes with special spin configurations
\cite{WBZ0601,FKM0703,Moore0978} on the edge under the protection of time
reversal symmetry. The experimental detection of such edge
modes\cite{KWB0766,HXQ0901,CAC0978} has attracted much attention and a lot of
efforts have been put into the exploration of new SPT phases.


Recently, we presented a systematic construction of SPT phases in bosonic
systems\cite{CGLW1172}, hence extending the understanding of SPT phases from
free fermion systems like topological insulators to systems with strong
interactions. We showed that there is a one-to-one correspondence between 2D
bosonic SPT phases with symmetry $G$ and elements in the third cohomology group
$\cH^3[G,U(1)]$. Moreover, we proved that\cite{CLW1141} due to the existence of
the special effective non-onsite symmetries on the edge of the constructed SPT
phases which are related to the nontrivial elements in $\cH^3[G,U(1)]$, the
edge states must be gapless as long as symmetry is not broken. However, it is
not clear what is the form of the gapless edge states, especially the experimentally more relevant low energy part. 


A low energy effective edge theory is desired because it could provide a simple
understanding of why the gapless edge is stable in these SPT phases. For
example, understanding of the low energy `helical' edge\cite{WBZ0601} in 2D
topological insulators enables us to see that some of the relevant gapping
terms are prohibited due to time reversal symmetry. Moreover, low energy
excitations are directly related to the response of the SPT phases to various
experimental probes, which has led to many proposals about detecting the exotic
properties of topological
insulators\cite{FK0807,FK0702,QWZ0608,QHZ0824,EMV0905,TYN0902}. Such an
understanding is hence also important for the experimental realization of
bosonic SPT phases.


In this paper, we study the low energy effective edge theory of the 2D bosonic
SPT phases with $\mathbb{Z}_N$ and $U(1)$ symmetry. We find that the gapless
states on the 1D edge is non-chiral, as it should be due of the lack of
intrinsic topological order\cite{W9505} in the system. The special feature of
the edge states lies in the way symmetry is realized. In particular, we find
that symmetry is realized chirally at low energy, i.e. in an inequivalent way
on the right and left movers. Because of the existence of this chiral symmetry,
the direct scattering between the left and right moving branches of the low
energy excitations is prohibited which provides protection to the gapless edge.

We would like to mention that people have used $U(1)\times U(1)$ Chern-Simons
theory\cite{L,LV12arXiv} and $SU(2)$ non-linear $\si$-model\cite{LW} to
construct the edge states of the $U(1)$ SPT phases. However, it is not clear whether
we have obtained the edge states for all of the $U(1)$ SPT phases using those
field theory approaches.  The construction presented in this paper has the
advantage of having a direction connection to the third cohomology group 
$\cH^3[U(1),U(1)]$.  So we are sure that we have obtained  the edge states for
all of the $U(1)$ SPT phases.

We would also like to point out that the chiral symmetry leads to a chiral response
of the system to externally coupled gauge field even though the edge state as a
whole is non-chiral. In particular, we find that all of the $U(1)$ SPT phases
have an even-integer quantized Hall conductance. 

\Ref{CLW1141,CGLW1172} show that, due to the short range
entanglement in SPT phases, the edge of the systems exists as a purely local 1D
system with a special non-onsite symmetry related to group cohomology. This
enables us to study the edge physics in 1D without worrying about the 2D bulk.
We will start with an exact diagonalization of the edge Hamiltonian in the
$\mathbb{Z}_2$ SPT phase constructed in \Ref{CLW1141}. Insights from this model
are then generalized to construct a 1D rotor model with different symmetries
realizing the edge states of all $\mathbb{Z}_N$ and $U(1)$ SPT phases. Some
useful formulas of the third group cohomology $\cH^3[G,U(1)]$ are reviewed in
appendix A\ref{Gcoh}.





\textit{Edge state of $\mathbb{Z}_2$ SPT phase} -- In \Ref{CLW1141} we presented an explicit construction of a nontrivial bosonic SPT phase with $\mathbb{Z}_2$ symmetry. The edge Hilbert space is identified as a local 1D spin $1/2$ chain. The spin chain satisfies a $\mathbb{Z}_2$ symmetry constraint given by
\be
U_2=\prod_i X_i \prod_{i} CZ_{i,i+1}
\ee
where $X$,$Y$ and $Z$ are the Pauli matrices and $CZ$ acts on two spins as
$CZ=\ket{00}\bra{00} + \ket{01}\bra{01} + \ket{10}\bra{10} - \ket{11}\bra{11}$. 
We showed in \Ref{CLW1141} that this non-onsite symmetry operator is related to the nontrivial element in the third cohomology group of $\mathbb{Z}_2$ and hence the edge state must be gapless if symmetry is not broken. Here we study one possible form of the edge Hamiltonian which satisfies this symmetry
\be
H_2=\sum_i X_i+Z_{i-1}X_iZ_{i+1}
\ee
This Hamiltonian is gapless because we can map this model to an $XY$ model. The mapping proceeds as follows: conjugate the Hamiltonian with $CZ$ operators on spin $2i-1$ and $2i$ and then change between $X$ and $Z$ basis on every $(2i-1)$th spin. The Hamiltonian then becomes
\be
H_2'=\sum_i X_{i-1}X_i + Z_{i-1}Z_i
\label{H_Z2_XY}
\ee
Therefore, the low energy effective theory of this model is that of a compactified boson field $\varphi(x)$ with Lagrangian density 
\be
\mathcal{L}=\frac{1}{2}\left[(\partial_t \varphi)^2 - v^2(\partial_x \varphi)^2 \right]
\label{Lboson}
\ee
This is a simple gapless state with both left and right movers and can be easily gapped out with a mass term such as the magnetic field in the $z$ direction $B_z(\sum_i Z_i)$. 
However, such a term is no longer allowed when the transformed $\mathbb{Z}_2$ symmetry operation is taken into account: 
\be
U_2'=\prod_{2i} CX_{2i,2i-1} \prod_{2i} Z_{2i-1}X_{2i} \prod_{2i} CX_{2i,2i+1}
\ee
where $CX_{i,j}$ acts on spin $i$ and $j$ as $CX = \ket{00}\bra{00} + \ket{01}\bra{01} + \ket{11}\bra{10} + \ket{10}\bra{11}$. This symmetry constraint prevents any term from gapping the Hamiltonian without breaking the symmetry. 

\begin{figure}[tb]
\begin{center}
\includegraphics[scale=0.4]{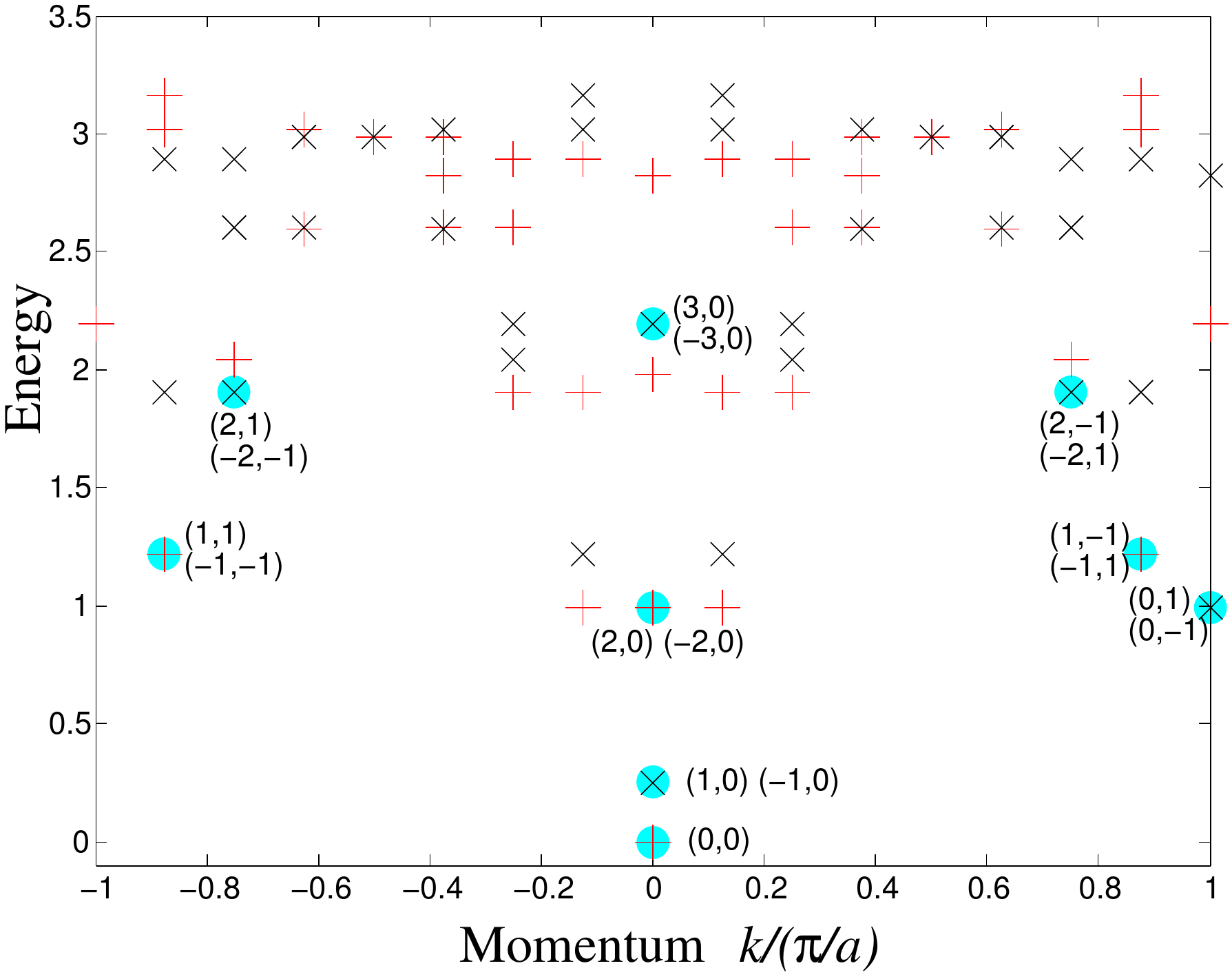}
\end{center}
\caption{Low energy states of $XY$ model $H_2'$(Eqn.(\ref{H_Z2_XY})). $x$-axis is lattice momentum
$k/(\pi/a)$, where $a$ is the lattice spacing. $y$-axis is energy with ground state
energy set to $0$ and first excited state energy normalized to $1/4$. $+$
represents positive $\mathbb{Z}_2$ quantum number and $\times$ represents negative
$\mathbb{Z}_2$ quantum number. Total boson number $l$ and winding number $m$
are labeled as $(l,m)$ for each primary field, represented by the shaded
$+$ or $\times$. States in the same conformal
tower have the same $l$ and $m$.
}
\label{fig:XY_Z2X}
\end{figure}

To see more clearly how this symmetry protects the gaplessness of the system, we study how it acts on the low energies modes. We perform an exact diagonalization of the $XY$ Hamiltonian Eqn.(\ref{H_Z2_XY}) for a system of $16$ spins and identify the free boson modes. Then we calculate the $\mathbb{Z}_2$ quantum number on these modes as shown in Fig. \ref{fig:XY_Z2X}. Note that $U_2'$ is not translational invariant and does not commute with the $U(1)$ symmetry of the $XY$ model $\prod_j e^{i\theta Y_j}$, therefore the free boson modes are not exact eigenstates of the $\mathbb{Z}_2$ symmetry. However, at low energy, the $\mathbb{Z}_2$ quantum number becomes exact as the system size gets larger and in Fig. \ref{fig:XY_Z2X} we plot the asymptotic $\mathbb{Z}_2$ quantum number of the low energy states.

From Fig. \ref{fig:XY_Z2X}, we can see that the $\mathbb{Z}_2$ quantum
number of each state only depends on the quantum number of the zero modes. The
zero modes are described by two integers: the total boson number $l$ and the
winding number $m$, and the $\mathbb{Z}_2$ symmetry at low energy acts as
$U_2' \sim (-)^{l+m}$.
From Fig. \ref{fig:XY_Z2X}, we also see that
the primary fields labeled by $(l,m)$ have the following
left- and right-scaling dimensions:
$(h_R,h_L)=( \frac{(l+2m)^2}{8}, \frac{(l-2m)^2}{8})$.

For the trivial $\mathbb{Z}_2$ SPT phase, the onsite $\mathbb{Z}_2$
transformation at low energy acts as $U_2' \sim (-)^{l}$, which is a non-chiral
action.  For the non-trivial $\mathbb{Z}_2$ SPT phase, we see that the
non-onsite $\mathbb{Z}_2$ transformation at low energy acts as $U_2' \sim
(-)^{l+m}$. We call such an $m$-dependent $U_2'$ a chiral symmetry operation.

From the chiral symmetry operation, we can have a simple (although not general) understanding of why some of the gap opening perturbations cannot appear in this edge theory. For example, the simplest mass term in the free boson theory $\int dx \cos(\varphi(x))$ contains a direct scattering term $\varphi_R(x)\varphi_L(x)$ between the left and right movers which carries a nontrivial quantum number under this $\mathbb{Z}_2$ symmetry and is hence not allowed. This result is consistent with that obtained by Levin $\&$ Gu\cite{LG12arXiv}.



\textit{Edge state of $\mathbb{Z}_N$ SPT phase} -- Understanding of how symmetry acts
chirally on the edge state of the $\mathbb{Z}_2$ SPT phase suggests that similar
situations might appear in other SPT phases as well. In this section we are
going to show that it is indeed the case for $\mathbb{Z}_N$ bosonic SPT phases. From the
group cohomology construction, we know that there are $N$ $\mathbb{Z}_N$-SPT phases
which form a $\mathbb{Z}_N$ group among themselves. We are going to construct
1D rotor models to realize the edge state in each SPT phase which satisfies
certain non-onsite symmetry related to the nontrivial elements in
$\cH^3[\mathbb{Z}_N,U(1)]$. From these models we can see explicitly how the symmetry
acts in a chiral way on the low energy states. Taking the limit of $N \to
\infty$ in $\mathbb{Z}_N$ will lead to the understanding of the edge states in $U(1)$
SPT phases which we discuss in the next section. Note that the choice of the
local Hilbert space on the edge, here a quantum rotor, is arbitrary and will
not affect the universal physics of the SPT phase as long as the effective
symmetry belongs to the same cohomology class. 

Consider a 1D chain of quantum rotors describe by $\{\varphi_i \}\in
(-\pi,\pi]$ with conjugate momentum $\{L_i\}$. The dynamics of the chain is
given by Hamiltonian
\be
H_r=\sum_i (L_i)^2 + V \cos(\varphi_i-\varphi_{i-1})
\label{Hr}
\ee
When $V>>1$, the system is in the gapless superfluid phase. At low energy,
$\varphi$ varies smoothly along the chain. The gapless low energy effective
theory is again described by a compactified boson field $\varphi(x)$ with
compactification radius $1$. The low energy excitations contain both left and
right moving bosons. 

The generator of the non-onsite $\mathbb{Z}_N$ symmetry related to the $M$th element ($M=0,...,N-1$) of the cohomology group, hence the $M$th SPT phase with $\mathbb{Z}_N$ symmetry, takes the following form in this rotor chain:
\be
U^{(M)}_N = \prod_{i} CP^{(M)}_{i,i+1} \prod_i e^{i2\pi L_i/N}
\label{UMN}
\ee
where $CP^{(M)}_{i,i+1}$ acts on two neighboring rotors and depends on $M$ as
\begin{equation*}
CP^{(M)}_{i,i+1} = \int d\varphi_i d\varphi_{i+1} e^{iM(\varphi_{i+1}-\varphi_i)_r/N} \ket{\varphi_i\varphi_{i+1}}\bra{\varphi_i\varphi_{i+1}}
\end{equation*}
Here we need to be careful with the phase factor
$e^{iM(\varphi_{i+1}-\varphi_i)/N}$ because it is not a single-valued function. We
confine $\varphi_{i+1}-\varphi_i$ to be within $(-\pi,\pi]$ and denote it as
$(\varphi_{i+1}-\varphi_i)_r$. Then $e^{iM(\varphi_{i+1}-\varphi_i)_r/N}$ becomes
single valued but also discontinuous when $\varphi_{i+1}-\varphi_i \sim \pm\pi$.
The discontinuity will not be a problem for us in the following discussion.
Note that it is important that $e^{iM(\varphi_{i+1}-\varphi_i)_r/N} \neq
e^{iM\varphi_{i+1}/N}/e^{iM\varphi_i/N}$, because otherwise the symmetry factors
into onsite operations and becomes trivial. We show in appendix C\ref{UMN_coh}
that $U^M_N$ indeed generates a $\mathbb{Z}_N$ symmetry. Moreover from its
matrix product unitary operator representation we find that the transformation
among the representing tensors are indeed related to the $M$th element in the
cohomology group $\cH^3[\mathbb{Z}_N,U(1)]$. Therefore, the 1D rotor model
represents one possible realization of the edge states in the corresponding SPT
phases. (The matrix product unitary operator formalism and its relation to
group cohomology was studied in \Ref{CLW1141} and we review the main results in
appendix B\ref{MPUO}).

The symmetry operator $U^{(M)}_N$ has a complicated form but its physical
meaning will become clear if we consider its action on the low energy states of
the rotor model in Eqn.(\ref{Hr}). First the $\prod_i e^{i2\pi L_i/N}$ part
rotates all rotors by the same angle $2\pi/N$, which can be equivalently
written as $e^{i2\pi L/N}$ with $L=\sum_i L_i$ being the total angular momentum
of the rotors. At low energy, $L$ is the total boson number with integer values
$l$. Moreover, at low energy $\varphi$ varies smoothly along the chain
therefore $(\varphi_{i+1}-\varphi_i)_r \sim \partial_x \varphi(x) dx$ and
$CP^{(M)}_{i,i+1}$ adds a phase factor to the differential change in $\varphi$
along the chain. Multiplied along the whole chain $\prod_{i} CP^{(M)}_{i,i+1}$
is equal to $e^{i2\pi M(\int dx \partial_x\varphi(x))/N}=e^{i2\pi Mm/N}$ where
$m$ is the winding number of the boson field $\varphi(x)$ along the chain. Put
together we find that the symmetry acts on the low energy modes as
\be
\label{UMN1}
U^{(M)}_N \sim e^{i2\pi(l+Mm)/N}
\ee
If $M$ is zero, this symmetry comes from a trivial SPT phase and $U^{(0)}_N$
depends only on $l$ which involves the left and right movers equally,as one can
see from right- and left-scaling dimensions $(h_R,h_L)=( \frac{(l+2m)^2}{8},
\frac{(l-2m)^2}{8})$. However, when $M$ is nonzero, this symmetry comes from a
nontrivial SPT phase and $U^{(M)}_N$ depends on $l+Mm$ which involves the left
and right mover in an unequal way. 
Put it differently, the symmetry on the edge
of nontrivial $\mathbb{Z}_N$ SPT phases acts chirally. 
In particular, when $M=2$, the symmetry will act only on the right movers.
Similar to the
discussion in the $\mathbb{Z}_2$ case, we can see that the chiral symmetry
protects the gaplessness of the edge by preventing direct scattering between
the left and right branches.

One may notice that $H_r$(Eqn.(\ref{Hr})) does not exactly commute with the symmetry $U^{(M)}_N$, but this will not be a problem for our discussions. We note that the potential term $\cos(\varphi_i-\varphi_{i-1})$ does commute with $U^{(M)}_N$. The kinetic term $(L_i)^2$ commute with the part that rotates $\varphi$ but not the phase factor $e^{iM(\varphi_{i+1}-\varphi_i)_r/N}$. However, at low energy, $(\varphi_{i+1}-\varphi_i) \to 0$, therefore this term becomes irrelevant locally and commutation between the Hamiltonian and the symmetry operator is restored. At high energy, in order for the Hamiltonian to satisfy the symmetry, we can change the kinetic term to $\sum^{N-1}_{k=0} \left(U^{(M)}_N\right)^k (L_i)^2 \left(U^{(M)}_N\right)^{-k}$. The high energy dynamics will be changed. However, because $V>>1$ and we know that the modified Hamiltonian does not break the $U(1)$ symmetry of the rotor model and the system cannot be gapped (due to the nontrivial cohomology class related to the symmetry), the system remains in the superfluid phase. The change in the kinetic term does not affect our discussion about low energy effective physics.



\textit{Edge state of $U(1)$ SPT phase} -- Taking the limit of $N \to \infty$, we can generalize our understanding of $\mathbb{Z}_N$ SPT phases to $U(1)$ SPT phases. As we show in this section, the chiral symmetry action on the low energy effective modes on the edge of the $U(1)$ SPT phases leads to a chiral response of the system to externally coupled $U(1)$ gauge field, even though the low energy edge state is non-chiral. We calculate explicitly the quantized Hall conductance in these SPT phases from the commutator of local density operators on the edge and find that they are quantized to even integer multiples of $\sigma_H=e^2/h$. In these SPT phases, a nonzero $U(1)$ Hall conductance exists despite a zero thermal Hall conductance.

>From group cohomology, we know that there are infinite 2D bosonic SPT
phases with $U(1)$ symmetry which form the integer group $\mathbb{Z}$ among
themselves. Generalizing the discussion in the previous section we find that
the low energy effective theory can be a $c=1$ free boson theory and the $U(1)$
symmetry acts on the low energy modes as $e^{i\alpha(l+Mm)}$, where $\alpha \in
[0,2\pi)$, $l$ is the total boson number, $m$ is the winding number and $M \in
\mathbb{Z}$ labels the $U(1)$ SPT phase. The local density operator of this
$U(1)$ charge is given by 
\begin{align}
\rho(x)=\Pi(x)+\frac{M}{2\pi}\partial_x \varphi(x),
\end{align}
with $\Pi(x)$ being the conjugate momentum of the boson field $\varphi(x)$, because the spatial integration of this density operator gives rise to the generator of the $U(1)$ symmetry $\int dx \rho(x) = l+Mm$. The commutator between local density operators is given by
\begin{align}
[\rho(x),\rho(x')] = -i\frac{2M}{2\pi}\delta'(x-x').
\end{align}
This term will give rise to a quantized Hall conductance along the edge when the system is coupled to an external $U(1)$ gauge field. Compared to the commutator between local density operators of a single chiral fermion 
\begin{align}
[\rho_{cf}(x),\rho_{cf}(x')] = -i\frac{1}{2\pi}\delta'(x-x'),
\end{align}
we see that the Hall conductance is quantized to even integer $2M$ multiples of $\sigma_H=e^2/h$.

As a consistency check we see that the quantized Hall conductance is a universal feature of the edge states in the bosonic $U(1)$ SPT phases and does not depend on the particular form the $U(1)$ symmetry is realized on the edge. Indeed, the $U(1)$ symmetry can be realized as $e^{i\alpha(Kl+K'm)}$, with arbitrary $K,K'\in \mathbb{Z}$. From the group cohomology calculation (reviewed in appendix B\ref{MPUO}) we find that it belongs to the cohomology class labeled by $M=KK'$. From the calculation of the commutator between local density operators, we see that the magnitude of the commutator is proportional also to $M=KK'$. Therefore, the Hall conductance depends only on the cohomology class--hence the SPT phase--the system is in and not on the details of the dynamics in the system.



\textit{Discussion} -- In this paper, we have construct the gapless edge
states for \emph{each} of the bosonic $\mathbb{Z}_N$ or $U(1)$ SPT phases in 2D.  We
show that those edge states are described by a $c=1$ non-chiral free boson
theory where the symmetry acts chirally on the low energy modes. The chiral
realization of the symmetry not only prevents some simple mass terms from
gapping out the system but also leads to a chiral response of the system to
external gauge fields. We demonstrate this by constructing explicit 1D lattice
models constrained by a non-onsite symmetry related to \emph{each} nontrivial
cohomology classes.  Our result indicates that the field theory approach based
on the $U(1)\times U(1)$ Chern-Simons theory\cite{L,LV12arXiv} and $SU(2)$
non-linear $\si$-model\cite{LW} indeed produce all of the $U(1)$ SPT phases.

We want to emphasize that although we have focused exclusively on the 1D edge, a
2D bulk having the 1D chain as its edge always exists and can be constructed by
treating a 1D ring as a single site and then putting the sites together. Note
that while the stability and chiral response of the edge in SPT phases are very
similar to that of the edge in quantum Hall systems, the underlying reason is
very different. The quantum Hall edge states are chiral in its own, which
remains gapless without the protection of any symmetry and leads to a nonzero
thermal Hall conductance.


Finally, we want to point out that the edge theory constructed in this paper is
only one possible form of realization. It is possible that other gapless
theories can be realized on the edge of SPT phases, for example with central
charge not equal to $1$. It would be interesting to understand in general what
kind of gapless theories are possible and what their universal features are. 

We would like to thank Zheng-Cheng Gu and Senthil Todadri for discussions. This work is supported by NSF DMR-1005541 and NSFC 11074140.


\section{Appendix A: The third group cohomology $\cH^3[G,U(1)]$ for symmetry $G$}
\label{Gcoh}

In this section, we will briefly describe the group cohomology theory. As we are focusing on 2D SPT phases, we will be interested in the third cohomology group.

For a group $G$, let $M$ be a G-module, which is an abelian group (with
multiplication operation) on which $G$ acts compatibly with the multiplication
operation (\ie the abelian group structure) on $M$:
\begin{align}
\label{gm}
 g\cdot (ab)=(g\cdot a)(g\cdot b),\ \ \ \ g\in G,\ \ \ \ a,b\in M.
\end{align}
For the cases studied in this paper, $M$ is simply the $U(1)$ group and $a$ an
$U(1)$ phase.  The multiplication operation $ab$ is the usual multiplication of
the $U(1)$ phases.  The group action is trivial: $g\cdot a=a$, $g\in G$, $a=\in
U(1)$.

Let $\om_n(g_1,...,g_n)$ be a function of $n$ group
elements whose value is in the G-module $M$. In other words, $\om_n:
G^n\to M$.  Let $\cC^n[G,M]=\{\om_n \}$ be the space of all such
functions.  
Note that $\cC^n[G,M]$ is an Abelian group
under the function multiplication 
$ \om''_n(g_1,...,g_n)= \om_n(g_1,...,g_n) \om'_n(g_1,...,g_n) $.
We define a map $d_n$ from $\cC^n[G,U(1)]$ to $\cC^{n+1}[G,U(1)]$:
\begin{align}
&\ \ \ \
(d_n \om_n) (g_1,...,g_{n+1})=
\nonumber\\
&
g_1\cdot \om_n (g_2,...,g_{n+1})
\om_n^{(-1)^{n+1}} (g_1,...,g_{n}) \times
\nonumber\\
&\ \ \ \ \
\prod_{i=1}^n
\om_n^{(-1)^i} (g_1,...,g_{i-1},g_ig_{i+1},g_{i+2},...g_{n+1})
\end{align}
Let
\begin{align}
 \cB^n[G,M]=\{ \om_n| \om_n=d_{n-1} \om_{n-1}|  \om_{n-1} \in \cC^{n-1}[G,M] \}
\end{align}
and
\begin{align}
 \cZ^n[G,M]=\{ \om_{n}|d_n \om_n=1,  \om_{n} \in \cC^{n}[G,M] \}
\end{align}
$\cB^n[G,M]$ and $\cZ^n[G,M]$ are also Abelian groups
which satisfy $\cB^n[G,M] \subset \cZ^n[G,M]$ where
$\cB^1[G,M]\equiv \{ 1\}$. $\cZ^n[G,M]$ is the group of $n$-cocycles and 
$\cB^n[G,M]$ is the group of $n$-coboundaries.
The $n$th cohomology group of $G$ is defined as
\begin{align}
 \cH^n[G,M]= \cZ^n[G,M] /\cB^n[G,M] 
\end{align}

In particular, when $n=3$, from
\begin{align}
&\ \ \ \ (d_3 \om_3)(g_1,g_2,g_3,g_4)
\nonumber\\
&= \frac{ \om_3(g_2,g_3,g_4) \om_3(g_1,g_2g_3,g_4)\om_3(g_1,g_2,g_3) }
{\om_3(g_1g_2,g_3,g_4)\om_3(g_1,g_2,g_3g_4)}
\end{align}
we see that
\begin{align}
& \cZ^3[G,U(1)]=\{  \om_3| 
\\
&\ \ \ \frac{ \om_3(g_2,g_3,g_4) \om_3(g_1,g_2g_3,g_4)\om_3(g_1,g_2,g_3) }
{\om_3(g_1g_2,g_3,g_4)\om_3(g_1,g_2,g_3g_4)}
=1
 \} .
\nonumber 
\end{align}
and
\begin{align}
& \cB^3[G,U(1)]=\{ \om_3| \om_3(g_1,g_2,g_3)=\frac{
\om_2(g_2,g_3) \om_2(g_1,g_2g_3)}{\om_2(g_1g_2,g_3)\om_2(g_1,g_2)}
 \},
\label{3coboundary}
\end{align}
which give us the third cohomology group
$\cH^3[G,U(1)]=\cZ^3[G,U(1)]/\cB^3[G,U(1)]$.


\section{Appendix B: Matrix Product Operator Representation of Symmetry}
\label{MPUO}

In \Ref{CLW1141} the symmetry operators on the edge of bosonic SPT phases were represented in the matrix product operator formalism from which their connection to group cohomology is revealed and the non-existence of gapped symmetric states was proved. In this section, we review the matrix product representation of the unitary symmetry operators and how the corresponding cocycle can be calculated from the tensors in the representation.

A matrix product operator acting on a 1D system is given by,\cite{MCP1012}
\be
O=\sum_{\{i_k\},\{i_k'\}}Tr(T^{i_1,i'_1}T^{i_2,i'_2}...T^{i_N,i'_N})|i'_1i'_2...i'_N\>\<i_1i_2...i_N|
\ee
where for fixed $i$ and $i'$, $T^{i,i'}$ is a matrix with index $\alpha$ and $\beta$. Here we are interested in symmetry transformations, therefore we restrict $O$ to be a unitary operator $U$. Using matrix product representation, $U$ does not have to be an onsite symmetry. $U$ is represented by a rank-four tensor $T^{i,i'}_{\alpha,\beta}$ on each site, where $i$ and $i'$ are input and output physical indices and $\alpha$, $\beta$ are inner indices.

\begin{figure}[ht]
\begin{center}
\includegraphics[scale=0.5]{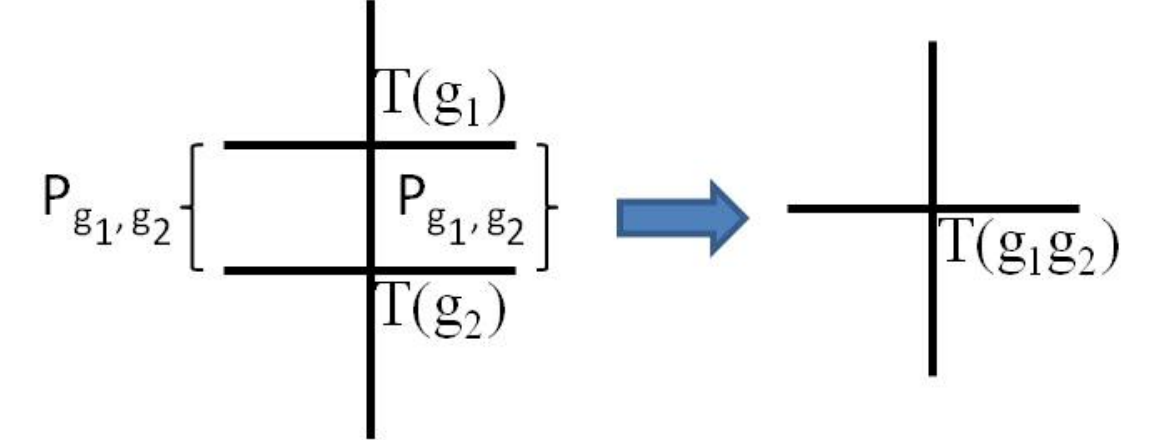}
\end{center}
\caption{Reduce combination of $T(g_2)$ and $T(g_1)$ into $T(g_1g_2)$. 
}
\label{P12}
\end{figure}

If $U(g)$'s form a representation of group $G$, then they satisfy $U(g_1)U(g_2)=U(g_1g_2)$. Correspondingly, the tensors $T(g_1)$ and $T(g_2)$ should have a combined action equivalent to $T(g_1g_2)$. However, the tensor $T(g_1,g_2)$ obtained by contracting the output physical index of $T(g_2)$ with the input physical index of $T(g_1)$, see Fig. \ref{P12}, is usually more redundant than $T(g_1g_2)$ and can only be reduced to $T(g_1g_2)$ if certain projection $P_{g_1,g_2}$ is applied to the inner indices (see Fig. \ref{P12}).

$P_{g_1,g_2}$ is only defined up to an arbitrary phase factor $e^{i\mu(g_1,g_2)}$. If the projection operator on the right side $P_{g_1,g_2}$ is changed by the phase factor $e^{i\mu(g_1,g_2)}$, the projection operator $P^{\dagger}_{g_1,g_2}$ on the left side is changed by phase factor $e^{-i\mu(g_1,g_2)}$. Therefore the total action of $P_{g_1,g_2}$ and $P^{\dagger}_{g_1,g_2}$ on $T(g_1,g_2)$ does not change and the reduction procedure illustrated in Fig.\ref{P12} still works. In the following discussion, we will assume that a particular choice of phase factors have been made for each $P_{g_1,g_2}$.

\begin{figure}[ht]
\begin{center}
\includegraphics[scale=0.5]{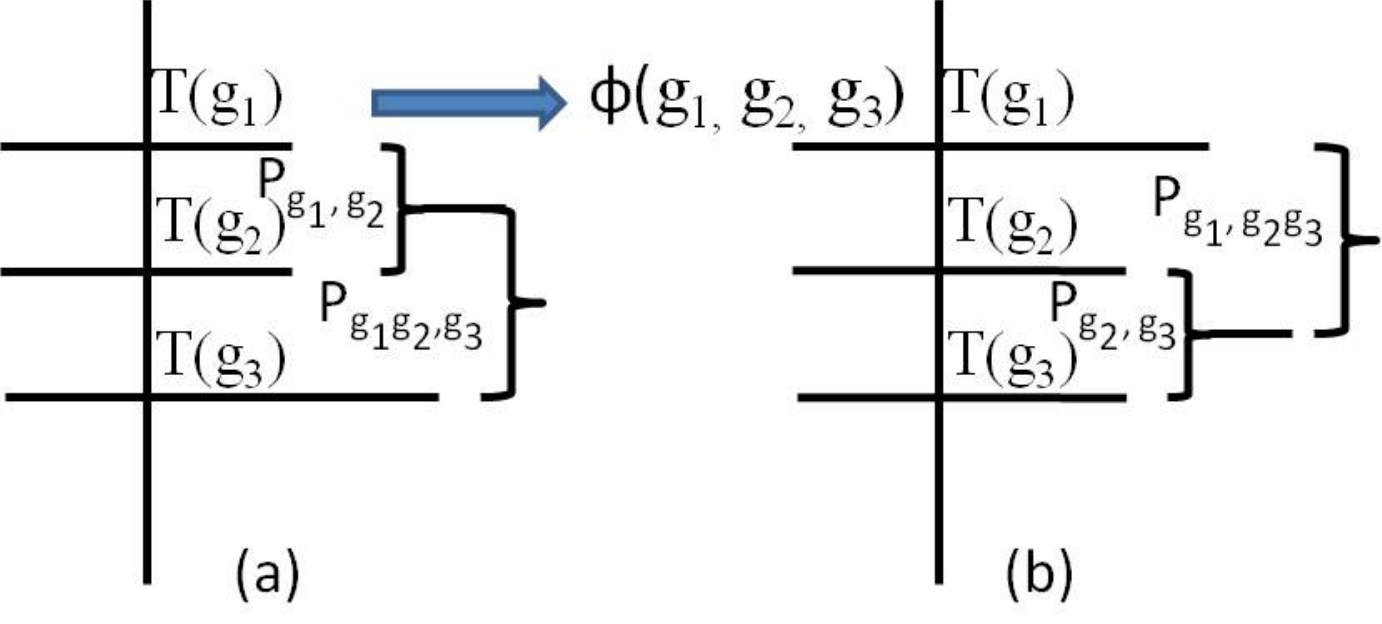}
\end{center}
\caption{Different ways to reduce combination of $T(g_3)$, $T(g_2)$ and $T(g_1)$ into $T(g_1g_2g_3)$. Only the right projection operators are shown. Their combined actions differ by a phase factor $\phi(g_1,g_2,g_3)$.
}
\label{P123}
\end{figure}

Nontrivial phase factors appear when we consider the combination of three symmetry tensors $T(g_1)$, $T(g_2)$ and $T(g_3)$. See Fig. \ref{P123}. 

There are two different ways to reduce the tensors. We can either first reduce the combination of $T(g_1)$, $T(g_2)$ and then combine $T(g_3)$ or first reduce the combination of $T(g_2)$,$T(g_3)$ and then combine $T(g_1)$. The two different ways should be equivalent. More specifically, they should be the same up to phase on the unique block of $T(g_1,g_2,g_3)$ which contributes to matrix contraction along the chain. Denote the projection onto the unique block of $T(g_1,g_2,g_3)$ as $Q_{g_1,g_2,g_3}$. We find that
\be
\begin{array}{l}
Q_{g_1,g_2,g_3}(I_3\otimes P_{g_1,g_2})P_{g_1g_2,g_3}= \\
\phi(g_1,g_2,g_3) Q_{g_1,g_2,g_3}(P_{g_2,g_3}\otimes I_1)P_{g_1,g_2g_3}
\end{array}
\ee
>From this we see that the reduction procedure is associative up to a phase factor $\phi(g_1,g_2,g_3)$. If we then consider the combination of four symmetry tensors in different orders, we can see that $\phi(g_1,g_2,g_3)$ forms a 3-cocycle of group $G$. That is, $\phi(g_1,g_2,g_3)$ satisfies
\be
\frac{ \phi(g_2,g_3,g_4) \phi(g_1,g_2g_3,g_4)\phi(g_1,g_2,g_3) }
{\phi(g_1g_2,g_3,g_4)\phi(g_1,g_2,g_3g_4)}
=1
\ee
The arbitrary phase factor of $P_{g_1,g_2}$ contributes a coboundary term to $\phi(g_1,g_2,g_3)$. That is, if we change the phase factor of $P_{g_1,g_2}$ by $\mu(g_1,g_2)$, then $\phi(g_1,g_2,g_3)$ is changed to
\be
\t \phi(g_1,g_2,g_3) = \phi(g_1,g_2,g_3)\frac{\mu(g_2,g_3)\mu(g_1,g_2g_3)}{\mu(g_1,g_2)\mu(g_1g_2,g_3)}
\ee
$\t \phi(g_1,g_2,g_3)$ still satisfies the cocycle condition and belongs to the same cohomology class as $\phi(g_1,g_2,g_3)$.


\section{Appendix C: Cohomology class of symmetry operator $U^{(M)}_N$ in Eqn.(\ref{UMN})}
\label{UMN_coh}

In this section, we discuss the property of the symmetry operator $U^{(M)}_N$ given in Eqn.(\ref{UMN}). First we show that $U^{(M)}_N$ indeed generates a $\mathbb{Z}_N$ symmetry. Next from its matrix product unitary operator representation we find that the transformation among the tensors are indeed related to the $M$th element in the cohomology group $\cH^3[\mathbb{Z}_N,U(1)]$. The calculation of cohomology class goes as described in the previous section. We repeat the definition of $U^{(M)}_N$ here
\be
U^{(M)}_N = \prod_{i} CP^{(M)}_{i,i+1} \prod_i e^{i2\pi L_i/N}
\label{UMN_a}
\ee
where $CP^{(M)}_{i,i+1}$ acts on two neighboring rotors and depends on $M$ as
\begin{equation*}
CP^{(M)}_{i,i+1} = \int d\varphi_i d\varphi_{i+1} e^{iM(\varphi_{i+1}-\varphi_i)_r/N} \ket{\varphi_i\varphi_{i+1}}\bra{\varphi_i\varphi_{i+1}}
\end{equation*}
Note that $(\varphi_{i+1}-\varphi_i)_r$ represents $\varphi_{i+1}-\varphi_i$ to be confined within $(-\pi,\pi]$. 

As $\prod_i e^{i2\pi L_i/N}$ rotates all the $\varphi_i$'s by the same angle and $\prod_{i} CP^{(M)}_{i,i+1}$ only depends on the difference between neighboring $\varphi$'s, the two parts in the symmetry operator commutes. Therefore
\begin{equation}
\left(U^{(M)}_N\right)^N = \prod_{i} \left(CP^{(M)}_{i,i+1}\right)^N \prod_i \left(e^{i2\pi L_i/N}\right)^N
\end{equation}
As $\prod_{i} \left(CP^{(M)}_{i,i+1}\right)^N = I$ and $\prod_i \left(e^{i2\pi L_i/N}\right)^N = e^{i2\pi L}=I$, $U^{(M)}_N$ indeed generators a $\mathbb{Z}_N$ symmetry on the 1D rotor system.

The matrix product representation of $U^{(M)}_N$ is given by
\begin{equation}
\begin{array}{l}
(T^{\varphi_0,\varphi_1})^{(M)}_N(1) = \delta(\varphi_1-(\varphi_0+\frac{2\pi}{N}))\times \\ 
 \int d\varphi_{\alpha} d\varphi_{\beta} \ket{\varphi_{\beta}}\bra{\varphi_{\alpha}} \delta(\varphi_{\beta}-\varphi_0)e^{iM(\varphi_{\alpha}-\varphi_0)_r/N}
\end{array}
\end{equation}
And the tensors representing $\left(U^{(M)}_N\right)^k$, $k\in \mathbb{Z}_N$ are given by
\begin{equation}
\begin{array}{l}
(T^{\varphi_0,\varphi_1})^{(M)}_N(k) = \delta(\varphi_1-(\varphi_0+\frac{2k\pi}{N}))\times \\ 
 \int d\varphi_{\alpha} d\varphi_{\beta} \ket{\varphi_{\beta}}\bra{\varphi_{\alpha}} \delta(\varphi_{\beta}-\varphi_0)e^{ikM(\varphi_{\alpha}-\varphi_0)_r/N}
\end{array}
\end{equation}Following the calculation described in the previous section, we find that the projection operation when combining $T^{(M)}_N(m_1)$ and $T^{(M)}_N(m_2)$ into $T^{(M)}_N((m_1+m_2)_N)$ is
\begin{equation}
\begin{array}{r}
P^{(M)}_N(m_1,m_2)=\int d\varphi_0\ket{m_2\frac{2\pi}{N}+\varphi_0}\ket{\varphi_0}\bra{\varphi_0} \times \\ 
e^{-iM\varphi_0(m_1+m_2-(m_1+m_2)_N)/N}
\end{array}
\end{equation}
where $(m_1+m_2)_N$ means addition modulo $N$.
When combining $T^{(M)}_N(m_1)$, $T^{(M)}_N(m_2)$ and $T^{(M)}_N(m_3)$, the phase angle in combining $m_1$ with $m_2$ first and then combining $(m_1+m_2)_N$ with $m_3$ is
\begin{equation}
\begin{array}{ll}
&M\varphi_0(-m_1-m_2+(m_1+m_2)_N -(m_1+m_2)_N- \\ 
&m_3+((m_1+m_2)_N+m_3)_N)/N\\ 
=&M\varphi_0(-(m_1+m_2+m_3)+(m_1+m_2+m_3)_N)/N
\end{array}
\end{equation}
the phase angle in combining $m_2$ with $m_3$ first and then combining $m_1$ with $(m_2+m_3)_N$ is
\begin{equation}
\begin{array}{ll}
&M\varphi_0(-m_2-m_3+(m_2+m_3)_N-m_1-\\ 
&(m_2+m_3)_N+ (m_1+(m_1+m_2)_N)_N)/N+ \\ 
&Mm_1\frac{2\pi}{N}(-m_2-m_3+(m_2+m_3)_N)/N \\ 
=&M\varphi_0(-(m_1+m_2+m_3)+(m_1+m_2+m_3)_N)/N + \\ 
&Mm_1\frac{2\pi}{N}(-m_2-m_3+(m_2+m_3)_N)/N
\end{array}
\end{equation}
Therefore, the phase difference is
\begin{equation}
\begin{array}{l}
\phi^{(M)}_N(m_1,m_2,m_3)  = \\ 
e^{iMm_1\frac{2\pi}{N}(-m_2-m_3+(m_2+m_3)_N)/N}
\end{array}
\end{equation}
We can check explicitly that $\phi^{(M)}_N(m_1,m_2,m_3)$ satisfies the cocycle condition
\begin{equation}
\begin{array}{l}
\frac{\phi^{(M)}_N(m_2,m_3,m_4)\phi^{(M)}_N(m_1,(m_2+m_3)_N,m_4)\phi^{(M)}_N(m_1,m_2,m_3)}{\phi^{(M)}_N((m_1+m_2)_N,m_3,m_4)\phi^{(M)}_N(m_1,m_2,(m_3+m_4)_N)}\\ 
=1
\end{array}
\end{equation}
Also we see that $\{\phi^{(M)}_N\}$, $M=0,...,N-1$, form a $\mathbb{Z}_N$ group generated by $\phi^{(1)}_N$. Therefore, the tensor $T^{(M)}_N$ corresponds to the $M$th element in the cohomology group $\cH^3[\mathbb{Z}_N,U(1)]$. 

Similar calculation holds for the $U(1)$ symmetry generated by $e^{i\alpha(Kl+K'm)}$, $K,K' \in \mathbb{Z}$. The cohomology class is labeled by $M=KK'$.


\section{Appendix D: Interpretation in terms of fermionization}

The free boson theory given in Eqn.(\ref{Lboson}) can be fermionized and the
low energy effective action of the symmetry discussed here can be reinterpreted
in terms of a free Dirac fermion. In particular, the fermionized theory has
Lagrangian density
\be
\mathcal{L}_f = \sum_{i=1,2} \psi^L_i(\partial_t + \partial_x)\psi^L_i + \psi^R_i(\partial_t - \partial_x)\psi^R_i
\label{Lfermion}
\ee
where $\psi_1$ and $\psi_2$ are two real fermions, out of which a complex
fermion can be defined $\Psi=\psi_1+i\psi_2$. Note that in order to have a
state to state correspondence between the boson and fermion theory, the fermion
theory contains both the periodic and anti-periodic sectors.

Since the $\mathbb{Z}_2$ symmetry in the nontrivial $\mathbb{Z}_2$ SPT phase only act on, say, the right moving sector,
one may naively guess that only $\psi_1^R$ change sign, while $\psi_2^R$,
$\psi_1^L$, and $\psi_2^L$ do not change under the  $\mathbb{Z}_2$
transformation: $ (\psi_1^R, \psi_2^R, \psi_1^L, \psi_2^L) \to (-\psi_1^R,
\psi_2^R, \psi_1^L, \psi_2^L) $.  In this case, the fermion mass term, such as
$(\psi_2^R)^\dagger \psi_2^L$, will be allowed by the $\mathbb{Z}_2$ symmetry.
Such a mass term will reduce the $c=1$ edge state to a $c=\frac 12$ edge state
without breaking the $\mathbb{Z}_2$ symmetry.  In the following, we will show
that the $\mathbb{Z}_2$ symmetry is actually realized in a different way.  The
$c=1$ edge state is stable if the $\mathbb{Z}_2$ symmetry is not broken.  So
the  $c=1$ edge state represents the minimal edge state for the $\mathbb{Z}_2$
(as well as the $\mathbb{Z}_N$ and $U(1)$) SPT phases.

The situation is best illustrated with explicit Jordan-Wigner transformation of the $XY$ model in Eqn.(\ref{H_Z2_XY}). Consider a system of size $N=4n$, $n \in \mathbb{Z}_+$. After the Jordan Wigner transformation
\begin{align}
\Psi_i=e^{i\pi\sum^{i-1}_{j=1}Z_j}(X_i+iY_i) \\ \nonumber
\Psi^{\dagger}_i=e^{i\pi\sum^{i-1}_{j=1}Z_j}(X_i-iY_i)
\end{align}
The Hamiltonian becomes
\begin{equation}
\begin{array}{l}
H=H_a+H_b \\ 
H_a=\sum^N_{i=1} (\Psi^{\dagger}_{i+1}\Psi_i+\Psi_i^{\dagger}\Psi_{i+1}) \\ 
H_b=-(P+1)(\Psi^{\dagger}_{1}\Psi_N+\Psi_N^{\dagger}\Psi_1)
\end{array}
\end{equation}
where $P=e^{i\pi\sum_{i=1}^N \Psi_j^{\dagger}\Psi_j}$ is the total fermion parity in the chain and $H_b$ is the boundary term which depends on $P$. Therefore, the fermion theory contains two sectors, one with an even number of fermions and therefore anti-periodic boundary condition and one with an odd number of fermions and periodic boundary condition. Without terms mixing the two sectors, we can solve the free fermion Hamiltonian in each sector separately. After Fourier transform, the Hamiltonian becomes
\be
H=\sum_k \cos\left(\frac{2\pi k}{N}\right) \Psi_k^{\dagger}\Psi_k
\ee
where $k$ takes value $0$, $1$, ..., $N-1$ in the periodic sector and $\frac{1}{2}$, $\frac{3}{2}$, ... $\frac{2N-1}{2}$ in the anti-periodic sector. The ground state in each sector has all the modes with energy $\leq 0$ filled. Note that with this filling the parity constraint in each sector is automatically satisfied. The ground state energy in the periodic sector is higher than in the anti-periodic sector and the difference is inverse proportional to system size $N$.

Now let's consider the effect of various perturbations on the system. 

The $(l,m)=(1,0)$ operator or the $(-1,0)$ operator in the boson theory (as
shown in Fig. \ref{fig:XY_Z2X}) corresponds to changing the boundary condition
of the Dirac fermion from periodic to anti-periodic. 
Such operators would totally gap out the edge states. However, from Eqn.
(\ref{UMN}) and Eqn. (\ref{UMN1}), we see that both operators carry nontrivial
quantum number in all 
$\mathbb{Z}_N$ (and $U(1)$) SPT phases, therefore it is forbidden by the
symmetry.

The $(l,m) = (2,0)$ operator in the boson theory corresponds
to the pair creation operator $\Psi^{\dagger}_L\Psi^{\dagger}_R$ in the fermion
theory. Its combination with the $(-2, 0)$ operator ($\Psi_R\Psi_L$
in the fermion theory) would gap out the system, but due
to the existence of the two sectors the ground state would be
two fold degenerate. To see this more explicitly, consider the $XY$ model again where 
the combination of $(l,m)=(2,0)$ and $(-2,0)$ operators can be realized with an anisotropy term
\be
H^{XY}_{(2,0)} = \gamma \sum_i X_{i-1}X_i - Z_{i-1}Z_i
\ee
Under Jordan Wigner transformation, it is mapped to the p-wave pairing term
\begin{equation}
\begin{array}{l}
H_{(2,0)}=H_{a,(2,0)}+H_{b,(2,0)} \\ 
H_{a,(2,0)}=\gamma \sum^N_{i=1} (\Psi^{\dagger}_{i+1}\Psi^{\dagger}_i+\Psi_i\Psi_{i+1}) \\ 
H_{b,(2,0)}=-\gamma(P+1)(\Psi^{\dagger}_{1}\Psi_N+\Psi_N^{\dagger}\Psi_1)
\end{array}
\end{equation}
Again, we have period boundary condition for $P=-1$ and anti-periodic boundary condition for $P=1$. After Fourier transform, the Hamiltonian at each pair of $k$ and $N-k$ is
\begin{equation}
\begin{array}{lll}
H_{k,N-k} &=& \cos\left(\frac{2\pi k}{N}\right)(\Psi_k^{\dagger}\Psi_k+\Psi_{N-k}^{\dagger}\Psi_{N-k}) + \\
& &i\gamma\sin\left(\frac{2\pi k}{N}\right) (-\Psi_k^{\dagger}\Psi^{\dagger}_{N-k}+\Psi_{N-k}\Psi_{k})
\end{array}
\end{equation}
The Bogoliubov modes changes smoothly with $\gamma$ and the ground state parity
remains invariant. The ground state energy is $\frac{1}{2}\sum_k
\left(1-(1-\gamma^2)\sin^2\left(\frac{2\pi k}{N}\right)\right)^{1/2}$ and
explicit calculation shows that the energy difference of the two sectors (with
$k = $ int. and $k=$ int. $+\frac12$) becomes exponentially small with nonzero
$\gamma$. Therefore, upon adding the $(l,m)=(2,0)$ and $(-2,0)$ terms, the
ground state becomes two fold degenerate. Such an operator does carry trivial
quantum number in the nontrivial $\mathbb{Z}_2$ SPT phase and renders the
gapless edge unstable. However, a two fold degeneracy would always be left over
in the ground states, indicating a spontaneous $\mathbb{Z}_2$ symmetry
breaking at the edge.

The $(0,1)$ operator in the boson theory corresponds to a scattering term
between the left and right moving fermions $\Psi^{\dagger}_L\Psi_R$. Its
combination with the $(0,-1)$ operator ($\Psi^{\dagger}_R\Psi_L$ in the fermion
theory) would gap out the system. Unlike the $(2,0)$ operator, there is no
degeneracy left in the ground state. In the $XY$ model, this corresponds to a staggered coupling constant
\be
H^{XY}_{(0,1)} = \gamma \sum_i(-1)^i \left(X_{i-1}X_i + Z_{i-1}Z_i\right)
\ee
Mapped to fermions, the Hamiltonian at $k$ and $k+\frac{N}{2}$ becomes
\begin{equation}
\begin{array}{lll}
H_{k,k+\frac{N}{2}} &=& \cos\left(\frac{2\pi k}{N}\right)(\Psi_k^{\dagger}\Psi_k-\Psi_{k+\frac{N}{2}}^{\dagger}\Psi_{k+\frac{N}{2}}) + \\  
& &i\gamma\sin\left(\frac{2\pi k}{N}\right) (-\Psi_k^{\dagger}\Psi_{k+\frac{N}{2}}+\Psi^{\dagger}_{k+\frac{N}{2}}\Psi_{k})
\end{array}
\end{equation}
For each pair of $k$ and ${k+\frac{N}{2}}$, 
there is one positive energy mode and one negative energy mode and we want to
fill the negative energy mode with a fermion to obtain to ground state.
For the anti-periodic sector, such a construction works
since there is a $N/2$ = even number of negative energy modes, and
the anti-periodic sector contains an even number of fermions.
However, for the periodic sector, such a construction fails
since there is a $N/2$ = even number of negative energy modes, and
the periodic sector must contain an odd number of fermions.
So we have to add an fermion to a  positive energy mode
(or have a hole in a negative energy mode), to have an odd number of
 fermions.
Therefore, the ground state in the periodic sector has a finite energy gap
above the anti-periodic one and the ground state of the whole system is
nondegenerate. However, because this term carries nontrivial quantum number in
any nontrivial $\mathbb{Z}_N$ (and $U(1)$) SPT phases, it is forbidden by the
symmetry.  For the trivial $\mathbb{Z}_2$ SPT phase, the $(0,\pm 1)$ operators
are $\mathbb{Z}_2$ symmetric operators, and can be added to the edge effective
Hamiltonian.  The presence of the $(0,\pm 1)$ operators will gap the edge state
and remove the ground state degeneracy.


\end{document}